\documentstyle[12pt]{article}
\def\sl2u1{SL(2,R)/U(1)}
\def\del{\partial}
\def\Ttilde{\tilde T}

\def\Vtilde{\tilde V}
\def\scF{{\cal F}}
\def\pifrac#1{\frac{#1}{2\pi i}}
\def\horspace#1{{\mbox{\hspace{0.5cm}#1\hspace{0.5cm}}}}
\def\ie{{\em i.e., }}

\begin{document}
\setlength{\oddsidemargin}{0cm}
\setlength{\baselineskip}{7mm}

\begin{titlepage}
    \begin{normalsize}
     \begin{flushright}
                 UT-Komaba/92-11 \\
                 November 1992
     \end{flushright}
    \end{normalsize}
    \begin{LARGE}
       \vspace{1cm}
       \begin{center}
         {Equivalence of BRST cohomologies for\\
          2-d black hole and $c=1$ Liouville theory} \\
       \end{center}
   \end{LARGE}

  \vspace{5mm}

\begin{center}
           Hiroshi I{\sc shikawa}
           \footnote{E-mail address:
              hishikawa@tkyvax.phys.s.u-tokyo.ac.jp}
                \ \  and \ \
           Mitsuhiro K{\sc ato}
           \footnote{E-mail address:
              katom@tkyvax.phys.s.u-tokyo.ac.jp} \\
      \vspace{4mm}
						{\it Institute of Physics, College of Arts and Sciences} \\
	     {\it University of Tokyo, Komaba}\\
      {\it Meguro-ku, Tokyo 153, Japan}\\
      \vspace{1cm}

    \begin{large} ABSTRACT \end{large}
\par
  \end{center}
\begin{quote}
\begin{normalsize}
\ \ \ \
We study the relation between the $SL(2,R)/U(1)$
black hole
and the $c=1$ Liouville theory.
A deformation, which interpolates the BRST operators of both models,
is explicitly constructed.
This interpolation is isomorphic, and the physical spectrum of the
black hole is equivalent to that of the $c=1$ model tensored by
a topological $U(1)/U(1)$ model.
Some implications of the deformation are discussed.
 \end{normalsize}
\end{quote}

\end{titlepage}
\vfil\eject

\section{Introduction}
\ \ \ \
Although string theory has many plausible properties,
it also has several difficulties which prevent us
to analyze it in detail.
One of them is
that we have no satisfactory formulation independent of
the background fields
in which strings propagate.
We have to specify the backgrounds before we construct a model
to examine.
The other is that there are no principle to single out one specific
vacuum from numerous candidates for a realistic model
which reproduces our real world in the low energy limit.
Towards the resolution of these problems, we have to
construct various kinds of models which have distinct backgrounds
and to understand their mutual relations.
But, this program is difficult to carry out, especially,
in the case of higher dimensional target space.
Higher dimension requires more degrees of freedom,
which make the analysis too complicated to handle.

Two-dimensional string theory, which is a theory with
a two-dimensional target space,
is appropriate for this aim, since it needs only
two degrees of freedom in order to describe a target space.
This model is a toy model from the point of view of
the unified string theory; our real world has, at least,
four dimensions.
However, it has many features peculiar to string theory,
and the study of it would give us deep insight to the nature
of string theory, specifically a clue to a background independent
formulation.
For this goal, it is necessary to have several models
and to uncover their relations.

In the case of two-dimensions,
their are two models examined in detail;
$c=1$ Liouville theory and $\sl2u1$ black hole proposed by Witten
\cite{WittenBH}.
Although many authors are discussing their relations
or correspondence with each other,
it seems that their is no satisfactory argument yet
\cite{BK,MS,DN,EKY}.

\vspace{2mm}
In this paper, we construct a one-parameter family of
the BRST operator
which interpolates between the $c=1$ model and the black hole.
Since this interpolation is isomorphic, we can extract the physical
spectrum of the black hole from that of the $c=1$ model.
Strictly speaking, what corresponds to the black hole
is not the $c=1$ model but the $c=1$ model tensored by a topological
degree of freedom.
Hence, the spectrum of the black hole is essentially that
of the $c=1$, which exhibits a close connection between these two.

In the next section, we prepare the $\sl2u1$ black hole
in the bosonized form,
which is the starting point of our analysis.
In Section 3, the interpolation of the BRST operator,
which can be considered as a deformation of the black hole
to the $c=1$ model,
is explicitly constructed.
The spectrum of the black hole is obtained by means of
this interpolation in Section 4.
Section 5 is devoted to some discussions and comments.

\section{$\sl2u1$ black hole}
\ \ \ \
The $\sl2u1$ black hole is a model which has the $\sl2u1$ manifold
as a target space. We describe this space using
the $SL(2,R)$ current
algebra
\begin{eqnarray}
J_{3}(z) \, J_{3}(w)    & \sim & \frac{-k/2}{(z - w)^{2}} \;\; ,\\
J_{3}(z) \, J^{\pm}(w)  & \sim & \frac{\pm 1}{z - w}
\, J^{\pm}(w) \;\; ,\\
J^{+}(z) \, J^{-}(w)    & \sim &
\frac{k}{(z-w)^2} - \frac{2}{z-w} \, J_{3}(w) \;\; .
\end{eqnarray}
$k$ is the level of this algebra.

This can be bosonized as follows;
\begin{eqnarray}
J_{3} & = & \sqrt{\frac{k}{2}} \, \del u \;\; ,\\
\label{sl2}
J^{\pm} & = &
i \, \left(
\sqrt{\frac{k^{'}}{2}}\, \del \phi \pm i \,
\sqrt{\frac{k}{2}}\,\del X \right)\, \exp \left(
\pm i\,\sqrt{\frac{2}{k}} (X + i\, u) \right) \;\;,
\end{eqnarray}
where $\phi$, $X$ and $u$ are free bosons with positive signature
satisfying, {\it e.g.\/},
$\phi(z) \phi(w) \sim - \log(z - w)$, and $k^{'} = k - 2$.
In this realization of the algebra, the energy-momentum tensor
by means of the Sugawara construction reads
\begin{eqnarray}
T_{SL(2,R)} & = & \frac{1}{k-2} \left(
\frac{1}{2} \left( J^{+}J^{-} + J^{-}J^{+} \right) - J_{3}J_{3}
\right) \nonumber \\
& = & -\frac{1}{2} (\del \phi)^{2} - \frac{1}{\sqrt{2k^{'}}}
\del^{2}\phi
-\frac{1}{2} (\del X)^{2} -\frac{1}{2} (\del u)^{2} \;\;.
\end{eqnarray}
The primary field $V_{j,m}$ which creates the state with spin $j$
and $J_{3} = m$ is realized as
\begin{equation}
\label{sl2_primary}
V_{j,m}(z) = \exp \left( \sqrt{\frac{2}{k^{'}}}\, j\, \phi(z)
\right)\,
\exp\left( i\,\sqrt{\frac{2}{k}}\, m (X(z) + i\, u(z)) \right) \;\; .
\end{equation}

In order to obtain the quotient space $\sl2u1$, we gauge away
a $U(1)$
degree of freedom generated by $J_{3}$. This gauging is performed by
a BRST operator \cite{KS}.
Since $J_{3}$ has a Schwinger term
in the operator product expansion
(OPE) with itself, we introduce a gauge boson $v$, satisfying
$v(z) v(w) \sim - \log (z-w)$,
to absorb this singularity.
Then the BRST operator for the $U(1)$ gauge fixing takes
the following form
\begin{equation}
Q_{U(1)} = \oint \pifrac{dz}\: \xi (J_{3} -
i\sqrt{\frac{k}{2}}\,
\del v) = \oint \pifrac{dz}\: \xi
\sqrt{\frac{k}{2}}(\del u - i \,\del v) \;\; .
\end{equation}
Here $\xi$, and its pair $\eta$, are fermionic ghosts with dimension
$h = 0$ and $1$, and obey $\xi(z) \eta(w) \sim 1/(z-w)$.
This BRST operator is nilpotent
\begin{equation}
Q_{U(1)}^{2} = 0 \;\; .
\end{equation}
Gauge fixing is now carried out by taking the BRST cohomology
as usual.
Since we introduced some new fields in this process,
the total energy-momentum tensor has additional contributions
\begin{equation}
T_{\sl2u1} = -\frac{1}{2} (\del \phi)^{2}
- \frac{1}{\sqrt{2k^{'}}}\del^{2}\phi
-\frac{1}{2} (\del X)^{2} -\frac{1}{2} (\del u)^{2}
-\frac{1}{2} (\del v)^{2} -\eta \, \del \xi \;\; .
\end{equation}
The central charge of this system reads
\begin{equation}
c_{\sl2u1} = 2 + \frac{6}{k - 2} \;\; .
\end{equation}
The $SL(2,R)$ primary fields (\ref{sl2_primary})
now turn into unphysical ones,
and we need to dress them properly in order to obtain
the physical operators.
The dressed or $\sl2u1$ primary fields $\Vtilde_{j,m}$
now can be written as
\begin{eqnarray}
\Vtilde_{j,m}(z) & = & V_{j,m}(z) \,
\exp\left( i\sqrt{\frac{2}{k}} m\, v(z) \right) \nonumber \\
& = & \exp\left( \sqrt{\frac{2}{k^{'}}} j \,\phi(z) \right) \,
\exp\left( i\sqrt{\frac{2}{k}} m (X(z) + i\, u(z) + v(z))\right)
\;\; .
\end{eqnarray}
It should be noted that zero modes of $X, u$ and $v$ appear only in
the combination of $X + i\,(u-i\,v)$.
This fact enables us to consider the Fock space of
this model as consisting of the states
with momentum in this combination.
We denote this space as $\scF_{\sl2u1}$.

Next, we couple this system to the world sheet gravity to construct
a string theory. Gauge fixing of the diffeomorphism invariance
requires $c_{\sl2u1} = 26$, which means $k = 9/4$.
We restrict ourselves to this critical case.
The BRST operator for the diffeomorphism invariance
is then defined as
\begin{equation}
Q_{diff} = \oint \pifrac{dz} \: (c\, T_{\sl2u1} + b c \,\del c),
\end{equation}
where $b$ and $c$ are the diffeomorphism ghosts with
$b(z)\, c(w) \sim 1/(z-w)$.
We define the total BRST operator as
\begin{equation}
\label{Q_BH}
Q_{BH} = Q_{diff} + Q_{U(1)}\;\; .
\end{equation}
Since $Q_{diff}$ is nilpotent and anti-commutes with $Q_{U(1)}$,
our BRST operator is nilpotent
\begin{equation}
Q_{BH}^{2} = 0\;\; .
\end{equation}

The Fock space of this model is now constructed by tensoring
that of $b$ and $c$ with $\scF_{\sl2u1}$.
We define the physical states of the black hole as
the $Q_{BH}$ cohomology in this Fock space.

\section{Deformation of the model}
\ \ \ \
Physical degrees of freedom in string theory are governed
by the BRST operator.
{}From this point of view, deforming a string theory is equivalent
to deforming a BRST operator keeping its nilpotency.
We apply this idea to the case of the black hole.

Let us consider the following deformation
\begin{equation}
\label{deformation}
Q(\alpha) = e^{\alpha\, R}\: Q_{BH}\: e^{-\alpha\, R}
\end{equation}
for a certain operator $R$ with dimension $0$ and
the ghost number $0$.
Note that this causes an automorphism in the operator algebra
and that $Q(\alpha)$ is necessarily nilpotent for
an arbitrary $\alpha$.
This operation is, therefore, a deformation of the model
in the sense above.
Our choice of the operator $R$ is
\begin{equation}
\label{R}
R = \oint \pifrac{dz}\: \frac{\sqrt{2}}{3} c \,\eta
(\del u + i\,\del v)\;\; .
\end{equation}
The explicit form of $Q(\alpha)$ then reads
\begin{equation}
\label{deformed_BRST}
Q(\alpha) = \oint \pifrac{dz} \left(
c \,\Ttilde(\alpha) + b c \,\del c \right) + Q_{U(1)} \;\; ,
\end{equation}
where
\begin{eqnarray}
\Ttilde(\alpha) & = &
-\frac{1}{2} (\del \phi)^{2} - \sqrt{2}\del^{2}\phi
-\frac{1}{2} (\del X)^{2}  \\
&   & \horspace{ }
+(1 - \alpha) \left(
-\frac{1}{2} (\del u)^{2} -\frac{1}{2} (\del v)^{2} -\eta \,\del \xi
+ \alpha \frac{\sqrt{2}}{3} \del c \,\eta (\del u + i \,\del v)
\right) \;\; . \nonumber
\end{eqnarray}
Note that $\Ttilde(\alpha) \neq
e^{\alpha\, R}\: T_{\sl2u1}\: e^{-\alpha\, R}$
and does not satisfy the Virasoro algebra in general.

The situation becomes very simple if we set $\alpha = 1$.
$\Ttilde(\alpha)$ then contains the first three terms only
which are nothing but the energy-momentum tensor of the $c=1$ model.
The BRST operator takes the form
\begin{equation}
Q(1)  = Q_{c=1} + Q_{U(1)} \;\; ,
\end{equation}
where
\begin{eqnarray}
Q_{c=1} & = & \oint\pifrac{dz}
\left(c\, T_{c=1}+ b c \,\del c \right)\;\; ,\\
T_{c=1}  & = & -\frac{1}{2} (\del \phi)^{2} - \sqrt{2}\,\del^{2}\phi
-\frac{1}{2} (\del X)^{2} \;\; .
\end{eqnarray}
$Q(1)$ splits into two anti-commuting pieces;
one operates on the $c=1$ sector and the other does on the rest
which spanned by $u, v, \xi$ and $\eta$.
In other words, we have, at $\alpha = 1$, a system of
the $c = 1$ model
tensored by additional degrees of freedom,
which completely decouple from the $c=1$ sector.
This $U(1)$ sector is a topological one,
a coset model $U(1)/U(1)$, and can be considered
to have no physical significance.
In fact, this sector, together with $Q_{U(1)}$ as a BRST operator,
realizes a twisted $N=2$ super conformal algebra,
which is a well-known example for the topological conformal field
theory.
Hence, our system at $\alpha = 1$ is essentially the $c=1$ model,
and the deformation generated by $R$ can be regarded as
an interpolation
between the $\sl2u1$ black hole and the $c=1$ model, which reveals
a close connection between these two.

\section{Physical spectrum of the black hole}
\ \ \ \
In this section, we study the physical spectrum of
the $\sl2u1$ black hole
making use of the interpolation obtained above.

The physical states are defined as the BRST cohomology.
Since the deformation (\ref{deformation}) causes an automorphism
in the operator algebra,
we need a cohomology for only one value of $\alpha$;
the results for the other values follow immediately.
Specifically, from a physical operator $V_{0}(z)$ of the black hole
($\alpha = 0$), we can construct that of the other value of $\alpha$
\begin{equation}
\label{automorphism}
V_{\alpha}(z) = e^{\alpha\, R}\: V_{0}(z) \: e^{- \alpha\, R} \;\; .
\end{equation}
This process can be reversed and allows us to determine
the physical spectrum of the black hole from that of the $c=1$ model.

The model at $\alpha =1$ is, however, not the $c=1$ model
but the $c=1$ model tensored by the $U(1)$ sector.
Therefore, we need at first to solve the cohomology of this system.
Note that we have a completely decoupling system, which means
that $Q_{c=1}$ and $Q_{U(1)}$ operate on its own sector
respectively and behave as a unit operator on another sector.
Hence, the cohomology $H_{Q(1)}^{*}$ of $Q(1)$ is nothing but
the tensor product of $Q_{c=1}$ and $Q_{U(1)}$
\begin{equation}
H_{Q(1)}^{*} = H_{c=1}^{*} \bigotimes H_{U(1)}^{*} \;\; ,
\end{equation}
which follows from the famous K\"{u}nneth formula \cite{Kunneth}.
Thus, what we need is only to determine
$H_{c=1}^{*}$ and $H_{U(1)}^{*}$ separately.

The cohomology $H_{c=1}^{*}$ of the $c=1$ system is known \cite{c=1}.
We shall show in Appendix that $H_{U(1)}^{*}$ consists of
two kinds of elements,
\begin{equation}
\label{H_u1}
H_{U(1)}^{*} =
\{ e^{k(u-i\, v)} \} \bigoplus \{ \xi\, e^{k(u-i\, v)} \}\;\; .
\end{equation}
In Section 2, however, we defined our Fock space as  containing
the zero modes of $u$ and $v$ in the combination of
$X + i\,(u-i\, v)$.
Hence, the momentum $k$ in (\ref{H_u1}) is fixed
if we specify the momentum of $X$.
This means that $H_{U(1)}^{*}$ is essentially two-dimensional
upon tensoring with $H_{c=1}^{*}$,
and the cohomology of $Q(1)$ is merely two copies of $H_{c=1}^{*}$.
As we mentioned above, the cohomology $H_{BH}^{*}$ of
the black hole is isomorphic to that of $Q(1)$.
Hence, our computation of $H_{Q(1)}$ leads us to
the following remarkable result;
\begin{equation}
H_{BH}^{*} \simeq H_{c=1}^{*} \bigoplus \xi_{0}\: H_{c=1}^{*}\;\; .
\end{equation}
The physical spectrum of the $\sl2u1$ black hole
coincides with that of the $c=1$ model except the zero mode
of the ghost.

By means of the different method from the present one,
Distler and Nelson \cite{DN} calculated the BRST cohomology
of the black hole to find out the extra states
besides the $c=1$ ones.
On comparing these two models, they adopted the following
correspondence between the label $j,m$ of the $SL(2,R)$
states and the $c=1$ momentum $p_{\phi}, p_{X}$;
\begin{equation}
\label{momentum}
p_{\phi} = \sqrt{8}\, j \;\;\;\;,\;\;\;\;
p_{X} = \frac{\sqrt{8}}{3}\, m \;\;.
\end{equation}
Let us examine this correspondence from our point of view.

In the $c=1$ model, the field with momentum, $p_{\phi}, p_{X}$,
reads
\begin{equation}
\label{c=1_field}
\exp\left( p_{\phi} \phi + i\, p_{X} X \right) \;\; .
\end{equation}
The black hole counterpart of this field
can be obtained making use of our automorphism
(\ref{automorphism}).
Because this does not
alter $\phi$ and $X$, the $c=1$ field (\ref{c=1_field})
is mapped into itself.
Comparing this expression (\ref{c=1_field}) with
the form (\ref{sl2_primary}) of the $SL(2,R)$ primary field,
we can conclude that
the field with the $c=1$ momentum $p_{\phi}, p_{X}$, is
turned into the field with the $SL(2,R)$ label $j,m$,
which are related by the formula (\ref{momentum})
adopted by Distler and Nelson.
Since the momenta of the physical states in our model
at $\alpha = 1$
are exactly the same as the $c=1$ model,
the spectrum of the black hole completely coincides with
that of the $c=1$ model; there are no extra states
in our formulation of the model.

\section{Discussion}
\ \ \ \
In the previous sections, we demonstrated that
the $c=1$ model can be constructed out of the $\sl2u1$ black hole
using an appropriate automorphism of the operator algebra.

A possible interpretation of the transformation is
a deformation of the target space imbedded in a higher
dimensional space.
The following facts support this observation.

First, the starting point of our analysis, the $\sl2u1$ black hole,
needs three dimensions in describing its target space,
which is a coset space $\sl2u1$ and can be considered
as a two-dimensional surface imbedded in a three-dimensional space.
The BRST operator $Q_{U(1)}$ for the $U(1)$ gauge fixing
restricts string propagation to this two-dimensional surface.
Second, our model at $\alpha = 1$ is not the $c=1$ model itself
but the $c=1$ model tensored by a topological degree of freedom
which can be regarded as a $U(1)/U(1)$ model.
It is, therefore, possible to look upon the $\alpha =1$ model
as a sort of coset model, $c=1 \bigotimes U(1)/U(1)$.
The BRST operator $Q_{c=1}$ for diffeomorphism
does not contain this $U(1)$ direction,
which suggests that the third or $U(1)$ coordinate
completely decouples from the target space,
\ie the target space of this model is a two-dimensional plane
in a three-dimensional space,
which is perpendicular to the $U(1)$ direction.
The BRST operator $Q_{U(1)}$ again serves as a constraint for
the target space.
Finally, it should be noted that
we fixed the Fock space of the model in deforming
the black hole by the operator $R$.
In other words, it was an automorphism of the operator algebra
that we made use of.
This seems to imply that
a three-dimensional space, in which the two-dimensional target space
is imbedded, is fixed in our deformation of the model,
while the target space, a two-dimensional surface,
is deformed within this three-dimensional space.
In order to make this picture more concrete one,
it would be necessary to examine our model
from more geometrical points of view, for example,
the sigma model approach.

\vspace{2mm}
As was pointed out in the last section,
the physical spectrum of the black hole in our formulation
is different from that of Distler and Nelson.
If our BRST operator $Q_{BH}$ (\ref{Q_BH}) annihilates
their extra states,
they are necessarily $Q_{BH}$-exact ones
because their momenta are off the $c=1$ grid
where the BRST cohomology is trivial.
It should be worth figuring out the meaning of their analysis.

\vspace{2mm}
In a recent paper, Eguchi, Kanno and Yang \cite{EKY}
also argued the equivalence of the $\sl2u1$ black hole to
the $c=1$ model.
They showed that the energy-momentum tensor of the black hole
can be rewritten as that of the $c=1$ up to BRST exact terms
under a suitable substitution of the fields,
and concluded that the physical spectrum of the black hole
is identical to that of the $c=1$.
It is, however, not the energy-momentum tensor
but the BRST operator that determines the spectrum
of the model in full.
We should discuss the correspondence of the BRST operators
rather than the energy-momentum tensors.
In \cite{EKY}, they adopted
a bosonized form of the $SL(2, R)$ current algebra
different from the present one (\ref{sl2}).
Their transformation of $\phi$ and $X$ and that of $b$ play
slightly different
roles.  The former is a part of the ``rotation'' that brings
the background
charge of four bosons together to one particular boson $\phi$,
and is only
needed for their form of the bosonization.
The latter is, on the other hand, relevant for (but gives
only a part of)
our transformation generated by
the $R$ (\ref{R}).\footnote[1]{Note that their fields
$\phi'$, $X'$ and $b'$ are not orthogonal to the field $\beta$.}

\vspace{2mm}
Lastly, we comment that the transformation of the BRST
charge of the $\sl2u1$ gauged WZW-model to that of the $c=1$
Liouville theory can be generalized to other models. Namely, for the
general $G/H$ gauged WZW-model coupled to 2-d gravity,  we have a
BRST charge in the form of $Q = Q_{diff} + Q_H$, where $Q_H$ is a
BRST charge for gauging a subgroup $H$ of $G$ and $Q_{diff}$ is that
for local diffeomorphism. Then we can construct the analogue of
$R$ in
eq.(\ref{R}) which transforms $Q$ into the direct sum $Q_{G/H} +
Q_H$, where $Q_{G/H}$ is the BRST charge of the GKO \cite{GKO}
type coset theory
coupled to 2-d gravity and $Q_H$ is that of the $H/H$ topological
theory.  An emphasis is again on the fact that the $Q_{G/H}$ and
$Q_H$ are completely decoupled.  Details will be reported elsewhere.

\vspace{0.5in}
\section*{Acknowledgements}
The work of M.K. is supported in part by
the Grant-in-Aid for Encouragement of Young Scientists (\# 04740139)
and by that for Scientific Research on
Priority Areas ``Infinite Analysis''
(\# 04245208) from the Ministry of Education, Science and Culture.

\newpage
\section*{Appendix  \ \ \ Cohomology of $Q_{U(1)}$}
\renewcommand{\theequation}{\Alph{section}.\arabic{equation}}
\setcounter{section}{1}
\setcounter{equation}{0}
\ \ \ \
We determine the cohomology of $Q_{U(1)}$,
\begin{equation}
Q_{U(1)} = \oint \pifrac{dz} \: \xi\:
\frac{3}{\sqrt{8}}(\del u - i\, \del v) \;\; .
\end{equation}
First, we define $P^{-}$ as
\begin{equation}
P^{-} = \oint \pifrac{dz} \: (\del u - i\, \del v) \;\; .
\end{equation}
Note that the cohomology is trivial unless $P^{-}$ vanishes,
because $P^{-}$ can be written as a BRST commutator
\begin{equation}
P^{-} \propto \{ Q_{U(1)}, \eta_{0} \} \;\; .
\end{equation}
One can show that the level counting operator $N$,
which counts the oscillator level of the states,
also can be written as a BRST exact one;
\begin{equation}
N = \{ Q_{U(1)}, K \} \;\;\mbox{, where}\;\;
K = \frac{3\, i}{\sqrt{2}}
\sum_{n \neq 0}\: \eta_{-n}(u_{n} + i\, v_{n}) \;\; .
\end{equation}
Therefore, the cohomology appears only in the level 0 sector.
Since the zero mode of $\xi$ does not vanish
on the $SL(2,C)$ vacuum, the cohomology of $Q_{U(1)}$
is spanned by two kinds of elements
\begin{equation}
H_{U(1)}^{*} =
\{ e^{k(u-i\, v)} \} \bigoplus \{ \xi\, e^{k(u-i\, v)} \}\;\; ,
\end{equation}
where $k$ runs all the value of the momenta.

\end{document}